\documentclass[12pt]{article}
\usepackage{epsfig}
\usepackage{graphicx}
\usepackage{amssymb}
\usepackage[tight]{subfigure}
\begin{document}

\begin{titlepage}

 \begin{center}
 {\LARGE \bf{Scalar overproduction in standard cosmology\\ 
  and\\ 
  \vspace*{.15in}
   predictivity of non--thermal dark matter}}
\end{center}

\vspace{1cm}

\begin{center}
  {\bf Oleg Lebedev}
\end{center}
  
\begin{center}
  \vspace*{0.15cm}
  \it{Department of Physics and Helsinki Institute of Physics,\\
  Gustaf H\"allstr\"omin katu 2a, FI-00014 Helsinki, Finland}\\
\end{center}
  
\vspace{2.5cm}

\begin{center} {\bf Abstract} \end{center}
\noindent    Stable scalars can be copiously produced in the Early Universe even if they have no coupling to other fields. We study production of
such scalars during and after (high scale) inflation, and obtain strong constraints on their mass scale. Quantum gravity-induced Planck-suppressed operators make an important impact 
on the abundance of dark relics.  Unless the corresponding Wilson coefficients are very small, they normally lead to  overproduction of dark states.
In the absence of a quantum gravity theory, such effects are uncontrollable, bringing into question predictivity of many non-thermal dark matter models.
These considerations may have non-trivial implications for string theory constructions, where scalar fields are abundant.

\end{titlepage}

  \tableofcontents

  \section{Introduction}
  
  Inflationary cosmology \cite{Starobinsky:1980te,Guth:1980zm,Linde:1981mu}
  addresses many conceptual  challenges of modern physics. It has become an integral part of the cosmological standard model, which assumes a period of inflation, followed
  by the inflaton oscillation epoch, reheating and a long period of radiation--dominated Universe evolution. These ingredients are sufficient to explain the observed structure of the Universe
  \cite{Mukhanov:2005sc}.

  In this work, we study  constraints on stable scalars in standard cosmology. Such fields  are produced in the Early Universe by various mechanisms, which can lead to overabundance of 
  dark relics. 
  Throughout this work, we make the following assumptions:
 \begin{itemize}
 \item  high scale inflation 
 \item existence  of a  stable scalar with mass below the inflationary Hubble rate and  the inflaton mass
\item the scalar is minimally coupled to gravity and has a very weak or no coupling to  other fields and itself
 \end{itemize}
  We consider both inflationary and postinflationary particle  production. These mechanisms are  efficient unless the scalar is very heavy. 
  The scalar field  is assumed to be very weakly coupled, possibly decoupled 
  from the observable and inflaton sectors, while its self-interaction  is  small enough such that 
 it does not reach thermal equilibrium. A special case of a dark relic of this type is non--thermal dark matter (DM).
 
  \vspace*{.2in}
\begin{figure}[!ht]
  \centering
  {\includegraphics[height=5.45em]{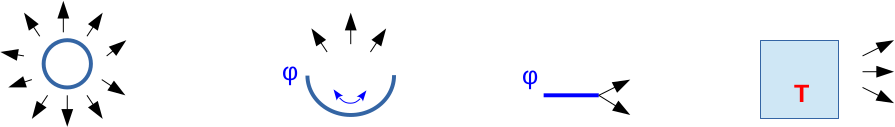}}
  \caption{ Typical dark relic production mechanisms: ({\it from left to right}) via inflation, inflaton oscillations, perturbative inflaton decay, thermal emission.  }
 \label{fig-intro}
\end{figure}
\vspace*{.2in}
 
  Very weakly interacting particles and  non--thermal dark matter, in particular, 
  have  $memory$ in the sense that their  abundance is additive and accumulates over different stages of the Universe evolution.
  As illustrated in Fig.\,\ref{fig-intro}, the most common production mechanisms are provided by (1) inflation, (2) inflaton oscillations, (3) perturbative inflaton decay,
  and (4) freeze--in type thermal emission. These are all very efficient even at tiny values of the couplings. If the scalar is completely decoupled
  from other fields, gravitational particle production still takes place during and after inflation, leading to strong constraints on its  mass scale.
  
  Quantum gravity effects play an important role in these considerations. Such effects are thought to generate (gauge-invariant) couplings between the different sectors of the model, typically
  in the form of higher dimensional Planck--suppressed operators. We show that these operators make an important impact on the abundance of dark relics, as long as the inflaton field value is large at the end of inflation.  This brings into question predictivity of non--thermal dark matter unless such operators are well under control. The latter is 
  only possible if  a  UV complete model of gravity is available.

The goal of this work is to obtain constraints on properties of dark feebly interacting relics 
assuming high scale inflation. These constraints turn out to be quite strong, especially taking into account quantum gravity effects.
While we focus mostly on scalar fields, some aspects of our study easily generalize to fermions (Sec.\,\ref{non-th-dm}).

  \section{Decoupled scalar production during inflation}

 Gravitational  particle production has been the subject of many research works
  \cite{Parker:1969au,Grib:1976pw,Ford:1986sy,Herring:2019hbe,Ford:2021syk}. Even if a given field has no couplings to other fields, it can be produced by gravity due to the space-time expansion.
 The latter creates the necessary non-adiabaticity if the Hubble rate is high enough.
 As a result, particles with sub-Hubble masses are abundantly produced, possibly causing cosmological problems \cite{Felder:1999wt,Kuzmin:1998kk}.\footnote{Some amount of heavy particles with masses above the Hubble rate is also
 generated \cite{Chung:1998zb,Chung:2018ayg}.} 
 
 Particle creation can also be formulated in terms of the field fluctuations characteristic of the de Sitter phase. This is done most conveniently using the corresponding equilibrium  probability distribution of
 Starobinsky and Yokoyama \cite{Starobinsky:1994bd}, which is the path we take in this work. 
 In Refs.\,\cite{Peebles:1999fz,Nurmi:2015ema,Markkanen:2018gcw}, this approach has been used to compute the generated dark matter abundance.

  Let us study inflationary scalar production following  Ref.\,\cite{Markkanen:2018gcw}.
  Consider a real scalar $s$ with the potential
  \begin{equation}
  V(s)= {1\over 2} m_s^2 \, s^2+ {1\over 4} \lambda_s \,s^4 \;,
  \end{equation}
which has no couplings to other fields apart from gravity. Suppose its self-coupling is weak   and it is effectively massless during inflation,
  \begin{equation}
  \lambda_s \ll 1 ~~,~~ m_s \ll H \;.
  \end{equation}
  In this case, the scalar experiences large quantum fluctuations induced by the Hubble expansion.
  The equilibrium distribution of the field is given by the probability density \cite{Starobinsky:1994bd}
  \begin{equation}
  P(s)\propto \exp\left[-8\pi^2 V(s)/(3H^4) \right] \;.
  \end{equation}
 This equilibrium is reached on the time scale $(\sqrt{\lambda_s}H)^{-1}$ \cite{Starobinsky:1994bd}. One can then distinguish two cases depending on the weakness of the coupling: the inflation period is long enough for the fluctuations to equilibrate and
 the inflation period is too short such that equilibrium never sets in.
 
   \subsection{Long inflation}
   
   In this case, one can read off the scalar fluctuation size from the equilibrium distribution.
    At the end of inflation, the variance of $s$ is given by
  \begin{equation}
  \langle s^2 \rangle \simeq 0.1 \times {H^2_{\rm end} \over {\sqrt \lambda_s}} \;,
  \end{equation}
  where $H_{\rm end }$ is the Hubble rate at the end of inflation. In other words, the scalar develops a condensate whose value is at least of the order of the Hubble rate and can be far greater than that if the self--coupling is very weak.\footnote{In the limit $\lambda_s=0$, the variance grows linearly in time $\propto H^3 t$.}

  After inflation ends, the condensate evolves through the following stages:
  \begin{itemize}
  \item it stays frozen as long as the Hubble rate is greater than the effective mass of $s$
  \item starts oscillating in an $s^4$ potential when the Hubble rate decreases to the effective mass of $s$
  \item oscillates in an $s^2$ potential when the effective mass of $s$ becomes comparable to the bare mass $m_s$, making it a non-relativistic dark relic
  \end{itemize}
  Let us now go through these stages in detail. For convenience, introduce the ``average'' field value 
  \begin{equation}
  \bar s \equiv \sqrt{  \langle s^2 \rangle} \;.
  \end{equation} 
  The effective mass of the scalar is given by
   \begin{equation}
 m^2_{\rm eff} = m_s^2 + 3 \lambda_s \bar s^2 \;.
  \end{equation} 
  Immediately after inflation, $m_{\rm eff} \ll H$, so that the potential can be neglected and the field is effectively ``frozen'', ${d\over dt} \bar s \simeq 0$.  
  For many purposes, the field can be treated as homogeneous, except for the 
  fluctuations generated by inflation. 
Soon after the end of inflation, the Universe becomes dominated by radiation. The precise nature of this transition (``reheating'') is unimportant for us.
In the instant reheating approximation, the reheating temperature $T_{\rm reh}$ is found via
  \begin{equation}
  3 H_{\rm end}^2 M_{\rm Pl}^2 = {g_* (T_{\rm reh}) \pi^2 \over 30} \, T_{\rm reh}^4 \;,
  \end{equation}
  where $g_*$ is the number of effective degrees of freedom at $T_{\rm reh}$.
  The resulting Hubble rate then decreases as $H \propto a^{-2}$ with the scale factor.
  
  The second stage in the evolution sets in when 
    \begin{equation}
 H_{\rm osc }^2 \sim m^2_{\rm eff}  \;.
 \label{Hosc}
  \end{equation} 
  We assume $m_s$ to be small enough such that $m_s^2 \ll  3 \lambda_s \bar s^2$ at this stage. The condensate starts oscillating in a quartic potential and behaves as radiation,
  \begin{equation}
  \bar s \propto a^{-1} \;.
  \end{equation}
  When the amplitude reduces further, the self-interaction term becomes comparable to the mass term in the potential,
  \begin{equation}
  {1\over 2} m_s^2 \bar s_{\rm dust}^2 \sim {1\over 4} \lambda_s \bar s^4_{\rm dust} ~~\Rightarrow ~~ \bar s_{\rm dust } \simeq {\sqrt{2 \over \lambda_s} \, m_s    } \;,
  \label{dust}
  \end{equation}
  which signifies the onset of the third phase in the evolution of $s$. After that, the field behaves as non--relativistic collisionless dust with the energy density scaling as $a^{-3}$. 
  In summary, we have the following stages in the evolution of $s$:
  \begin{equation}
\bar s_{\rm end}  ~{\stackrel{ a^0} { \longrightarrow   }  }   ~ \bar s_{\rm osc}    ~
 \stackrel{  a^{-1}} \longrightarrow \bar s_{\rm dust} \;,
\end{equation}
  
  The  number density in the ``dust'' phase is given by 
  \begin{equation}
  n (a)  = {\rho (a) \over m_s} \simeq {m_s^3\over \lambda_s } \, {a_{\rm dust}^3 \over a^3} \;.
  \end{equation}
 The abundance of the $s$  quanta is conveniently expressed in terms of the conserved quantity
    \begin{equation}
 Y= {n \over s_{\rm SM}} ~~,~~ s_{\rm SM} = {2\pi^2 \over 45}\, g_{*s} T^3 \;,
 \label{Y-def}
  \end{equation}
where $ s_{\rm SM}$ is the entropy density of the SM thermal bath at temperature $T$ 
     and $g_{*s}$ is the effective number of degrees of freedom contributing to the entropy.
     $Y$ can, for instance, be evaluated at the onset of the non--relativistic period characterized by temperature $T_{\rm dust}$. At  this stage, $n \simeq m_s^3/ \lambda_s$, while $s_{\rm SM}(T_{\rm dust})$
     is obtained from   $s_{\rm SM}(T_{\rm reh})$ by entropy conservation. The relevant scale factors $a_{\rm osc} = a_{\rm end}\, \sqrt{H_{\rm end} / H_{\rm osc}}$ and $a_{\rm dust}$ can be computed from
     (\ref{Hosc}) and (\ref{dust}), respectively. 
     
 Requiring that the abundance of $s$  not exceed that of dark matter,
 \begin{equation}
 Y \leq 4.4 \times 10^{-10} \;{{\rm GeV}\over m_s} \;,
 \label{Y-constraint}
 \end{equation}
  we obtain the constraint \cite{Markkanen:2018gcw}
  \begin{equation}
m_s\, \lambda_s^{-5/8} \lesssim 10^{-7}\, \left( {M_{\rm Pl}\over H_{\rm end}}\right)^{3/2} \,{\rm GeV} \;.
\label{ms-infl}
 \end{equation}
  For a typical $H_{\rm end} \sim 10^{14}$ GeV, the right hand side yields a number close to 1 GeV. Since $\lambda_s \ll 1$ for a feebly interacting scalar, its mass has to be far below the GeV scale,
  $
  m_s \ll {\rm GeV} \;,
 $
  otherwise the resulting Universe would be too dark.

  The above derivation assumes that there was no appreciable period of matter domination, which is the case, for instance, in the $\phi^4$ local inflaton potential.
  The presence of such a stage dilutes the abundance of the dark relic thereby relaxing the constraint. 
  Assuming that the Universe undergoes an extended period of matter domination  following  inflation,
   \begin{equation}
H_{\rm end} ~ {\stackrel{ {\scriptstyle{a^{-3/2}}}} { \longrightarrow   }  } ~   H_{\rm reh}   
   \;,
\end{equation}
   such that
  reheating happens after $\bar s$ starts oscillating,
  we get
    \begin{equation}
m_s\, \lambda_s^{-3/4} \lesssim 10^{-7}\,   \Delta_{\rm NR} \,\left( {M_{\rm Pl}\over H_{\rm end}}\right)^{3/2} 
  \,{\rm GeV} \;,
  \label{m-lambda-NR}
 \end{equation}
  where $\Delta_{\rm NR}$ characterizes the duration of the non--relativistic expansion period,
   \begin{equation}
 \Delta_{\rm NR} \equiv
\left( {H_{\rm end}\over H_{\rm reh}}\right)^{1/2} \;.
 \end{equation}
 We note that the power of $\lambda_s$ changes only slightly compared to that in the instant reheating case, while the right--hand side of (\ref{m-lambda-NR}) receives a potentially large factor 
  $\Delta_{\rm NR} = T_{\rm inst-reh}/T_{\rm reh} \gg 1$, where $T_{\rm inst-reh}$ is the reheating temperature assuming instant reheating. For example, in a local $\phi^2$ inflaton potential, this 
  ratio can be very large if the inflaton decays slowly. For $H_{\rm end} \sim 10^{14}$ GeV, the allowed scalar mass is bounded roughly by $ \Delta_{\rm NR}$,
\begin{equation}
  m_s \ll  \Delta_{\rm NR}  ~ {\rm GeV}\;.
  \label{infl-bound}
  \end{equation}
We note that, in typical  $\phi^2$  models, $ \Delta_{\rm NR}$  ranges from 1 to about $10^5$, while in the extreme case of an MeV-scale  reheating temperature \cite{Hannestad:2004px}, it can be as large as $10^{18}$.
  On the other hand, in the $\phi^4$ case, $ \Delta_{\rm NR}=1$.\footnote{ It is assumed  that, in the $\phi^4$ case, the inflaton mass can be neglected at the relevant energy scales. 
  This may not apply to models with  very low reheating temperatures  \cite{Lebedev:2021tas}, in which case an analog of $\Delta_{\rm NR} $ would have to be introduced.}

      \begin{figure}[h!] 
\centering{
\includegraphics[scale=0.33]{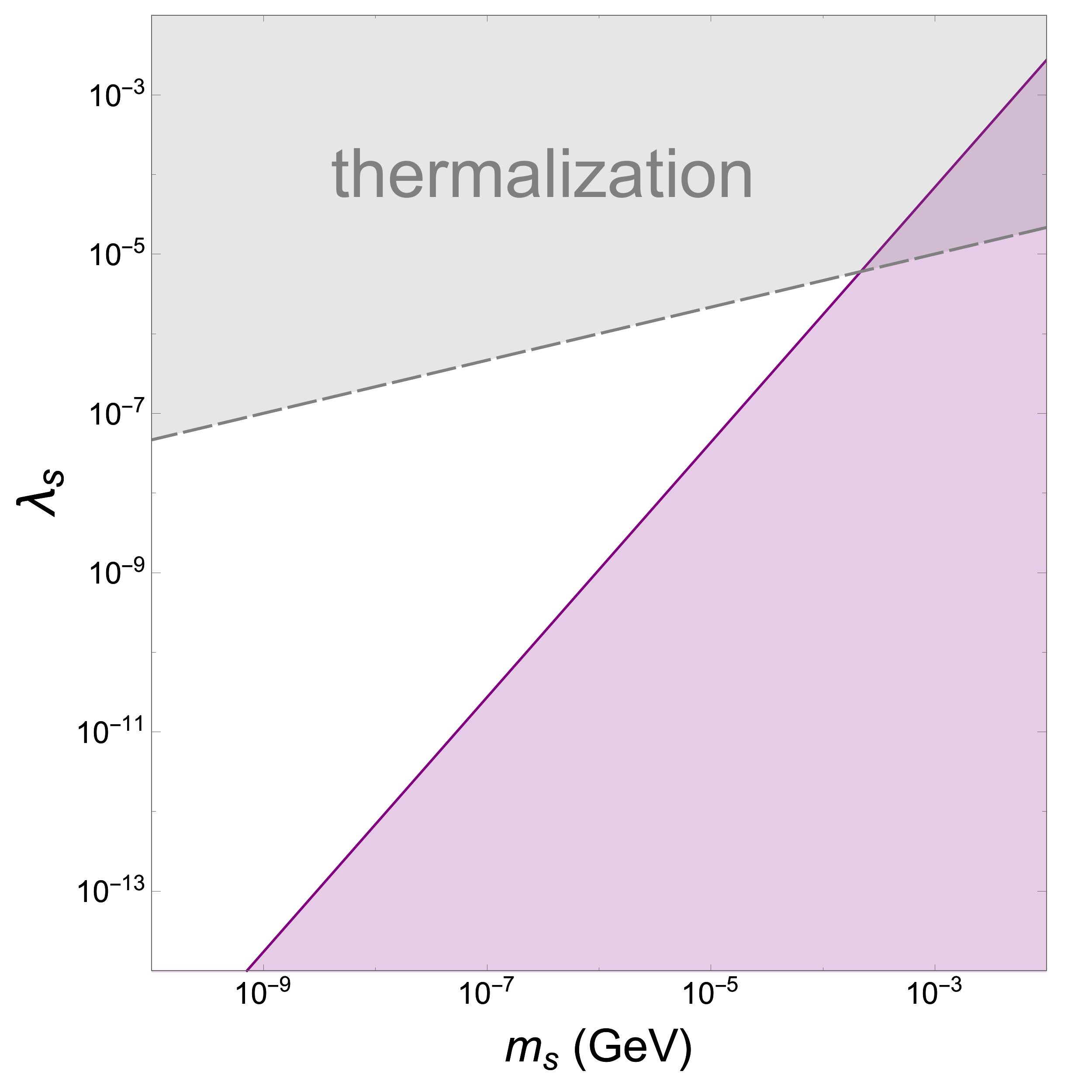}
}
\caption{ \label{m-lambda}
Scalar mass vs  self--coupling constraint for the equilibrium distribution of scalar fluctuations.
The darker shaded region is excluded by scalar overabundance  assuming a $\phi^4$ local inflaton potential or instant reheating with the Hubble rate at the end of inflation $H_{\rm end} \sim 10^{14}$ GeV. 
The existence of the matter dominated phase ($\phi^2$ local inflaton potential) relaxes the corresponding constraint on $m_s$ by a factor of $\Delta_{\rm NR}$. In the lighter shaded region, the non--thermalization assumption is violated.}
\end{figure}

  The excluded parameter space in the case of instant reheating  is shown in Fig.\,\ref{m-lambda}. The region with substantial couplings is inconsistent with our non--thermalization assumption.
  The thermalization constraint \cite{Arcadi:2019oxh} can be approximated by 
  \begin{equation}
  \lambda_s \simeq 10^{-4} \,(m_s/{\rm GeV})^{1/3} \; , 
  \end{equation}
  accounting for 
    the difference in the definition of $\lambda_s$. For smaller $\lambda_s$, the reaction rate $ss\rightarrow ssss$ is below the Hubble rate at any temperature such that thermal equilibrium cannot be reached.\footnote{This assumes that the dark sector is not much hotter than the observable sector.} The darker region is excluded by overabundance of $s$ assuming instant reheating with $H_{\rm end} \sim 10^{14}$ GeV, while for a general case the bound on $m_s$ relaxes by a factor of $ \Delta_{\rm NR}$ (see Eq.\,\ref{m-lambda-NR}).

  Further constraints apply if $s$ constitutes all of dark matter. In particular, the fluctuations in the energy density of $s$ are not correlated with those of the inflaton, hence there are significant isocurvature bounds
  \cite{Markkanen:2018gcw}. Also, for very light $s$,  there are strong bounds from structure formation (see, e.g.\,\cite{Garcia:2022vwm}). Here, we take a conservative view and treat $s$ as a stable relic whose energy density can be below that of dark matter.

  \subsection{Short inflation}
  
  The inflation period is allowed to be short enough such that it produces the required 60 $e$-folds, but the scalar fluctuations do not reach the equilibrium distribution before the end of inflation. 
  In this case, the analysis is model-dependent and one may simply use an estimate (see, e.g.\, \cite{Felder:1999wt})
  \begin{equation}
  \bar s_{\rm end} \sim H_{\rm end } \;.
  \end{equation}
  The rest of the calculation remains the same. One finds that in the $\phi^4$ or instant reheating case, the relic abundance $Y$ scales as $\bar s_{\rm end}^{3/2} \,\lambda_s^{-1/4}$, so
  the constraint reads
    \begin{equation}
m_s\, \lambda_s^{-1/4} \lesssim 10^{-8}\, \left( {M_{\rm Pl}\over H_{\rm end}}\right)^{3/2} \,{\rm GeV} \;.
 \end{equation}
  If a sufficiently long  non-relativistic expansion period takes place, $Y \propto \bar s_{\rm end} \, \lambda_s^{-1/2}$, and we get
    \begin{equation}
m_s\, \lambda_s^{-1/2} \lesssim 10^{-7}\,   \Delta_{\rm NR} \,\left( {M_{\rm Pl}\over H_{\rm end}}\right)^{3/2} 
  \,{\rm GeV} \;.
 \end{equation}
  Note  that the constraints on $m_s$ get progressively stronger for weaker couplings, as before. This is due to the fact that the number density of the non-relativistic scalars  at the ``dust'' epoch  scales as
  $m_s^3/\lambda_s$, which makes them more abundant for smaller $\lambda_s$.

  We conclude that  feebly-interacting stable scalars 
   are efficiently produced during inflation.{\footnote{Analogous considerations apply to particle production during the radiation-dominated phase of the Universe expansion, although the resulting bounds are 
much weaker, e.g.  $m_s \lesssim 10^8$ GeV  \cite{Kuzmin:1998kk}.}}
    Such scalars are overabundant  unless their mass scale is far below a GeV, depending on the self--coupling,  or the Universe  has undergone a long period of non-relativistic expansion.    
  
   These constraints can be evaded if the scalars have a tangible coupling to the inflaton, which creates an effective mass term during inflation. A similar effect is achieved with 
   a non--minimal scalar coupling to gravity. However, such couplings lead to very efficient  particle production in the inflaton oscillation epoch, which therefore only delays the problem. The corresponding 
   bounds will be studied in the subsequent sections.

    \section{Decoupled scalar production after inflation}
    \label{Ema-effect}

    After the end of inflation, the inflaton field $\phi$ experiences coherent oscillations around the potential minimum. This leads to 
    an oscillating Hubble rate and, thus, 
    particle production even in the absence of a direct coupling
    between  $\phi$ and the scalar in question \cite{Ema:2015dka,Ema:2016hlw}. The resulting scalar abundance can readily violate the observational bounds unless appropriate constraints are satisfied.

    Let us consider in detail this production mechanism following Ref.\,\cite{Ema:2015dka}.

    \subsection{Background dynamics}
    
    Consider the dynamics of a homogeneous inflaton field $\phi$. The Friedmann equation and the equation of motion for $\phi$ read
    \begin{eqnarray}
    && 3 M_{\rm Pl}^2 H^2 = \rho_\phi = {1\over 2} \dot \phi^2 + V \;,\\
    && \ddot \phi + 3 H \dot \phi + V^\prime_\phi =0 \;.
    \end{eqnarray}
    These yield 
    \begin{equation}
    \dot \rho_\phi +3H \dot \phi^2 =0\;,
    \label{dot-rho}
      \end{equation}
    which shows that $\rho_\phi$ decreases in time and is expected to oscillate when $\phi$ oscillates.
    Consider the regime $m_\phi^{\rm eff} \gg H$, where $m_\phi^{\rm eff}$ is the effective inflaton mass determining the oscillation frequency.  In this case, $\rho_\phi$ is approximately conserved on
    short time scales between $1/ m_\phi^{\rm eff}$ and $1/H$, and one may use the virial theorem stating that for 
     \begin{equation}
  V \propto \phi^n  \;,
      \end{equation}
$\langle \dot \phi^2 \rangle = n \langle V \rangle$, where the averaging is performed over multiple inflaton oscillations. Then, (\ref{dot-rho}) implies
\begin{equation}
\langle \dot \rho_\phi \rangle +{6n \over n+2} \langle H\rangle \langle \rho_\phi \rangle =0 \;,
\end{equation}
  which is solved by $  \langle \rho_\phi \rangle\propto \langle a(t)\rangle^{- {6n\over n+2}} $ such that $\langle H \rangle ={n+2\over 3nt}$.  
    
  As Eq.\,\ref{dot-rho} suggests, both $\rho_\phi$ and $H$ oscillate around their average values. The amplitude of these oscillations is small, yet they are of high frequency.
      Let us express $\rho_\phi = \langle \rho_\phi \rangle + \delta \rho_\phi$ and 
    $H = \langle H\rangle + \delta H$, where the perturbations are the oscillating parts.
    The explicit  form of $\delta H$ can be determined as follows. Using $\dot H = \dot \rho_\phi /( 2\sqrt{3} M_{\rm Pl} \sqrt{\rho_\phi})$ along with (\ref{dot-rho}) and 
    $ \langle \dot H \rangle =- {n+2\over 3nt^2}$, we get
    \begin{equation}
    \delta \dot H = {3n \over n+2 }  \langle  H \rangle^2 - {\dot \phi^2 \over 2 M_{\rm Pl}^2 } \;.
    \end{equation}
    To linear order in $\delta H$, $\langle  H \rangle^2$ can also be written as $ \langle  H \rangle^2 = {1\over 3M_{\rm Pl}^2} \left(   {1\over 2} \dot \phi^2 +V   \right) 
    -2 \langle  H \rangle \delta H$, which yields
     \begin{equation}
    \delta \dot H  +   {6n \over n+2 }       \langle  H \rangle \delta H     =   - {1\over  (n+2)  M_{\rm Pl}^2} \left(  {\dot \phi^2   } -nV\right) \;.
    \end{equation}
    By virtue of the  $\phi$ equation of motion  and $V^\prime_\phi \phi = nV$, the term ${\dot \phi^2   } -nV$ can be written as $(d/dt + 3H) \phi \dot \phi$. 
    Now, recall that the oscillations are of high frequency determined by $m^{\rm eff}_\phi \gg H$. Thus,  $\delta \dot H \gg H \delta H$ and $d/dt \, (\phi \dot \phi) \gg H \phi \dot \phi$. Neglecting the Hubble--suppressed terms, we get  $\delta \dot H \simeq -{1\over (n+2) M_{\rm Pl}^2} \,{d\over dt}\, \phi \dot \phi$, hence
    \begin{equation}
    \delta  H \simeq -{1\over (n+2) M_{\rm Pl}^2} \, \phi \dot \phi \;.
    \end{equation} 
    For $n=2$, this result can easily be verified by an explicit calculation \cite{Mukhanov:2005sc}. Since $H \sim {\cal O} (\dot \phi /M_{\rm Pl})$, the oscillating correction is suppressed by the Planck scale, 
    $\delta H / H \sim \phi / M_{\rm Pl} \ll 1$. 
    
    The above  equation allows us to determine  the oscillating part of the scale factor. Using the explicit form of $\langle H \rangle$ and integrating $\dot a /a = H = \langle H \rangle + \delta H$, we get
    (to linear order in $M_{\rm Pl}^{-2}$)
    \begin{equation}
    a(t) \simeq \langle a(t) \rangle \, \left[     1- {1\over 2(n+2)} \, {\phi^2 -\langle \phi^2 \rangle \over M_{\rm Pl}^2 }    \right] \;,
    \label{a-osc}
    \end{equation}
    where we have used the boundary condition $a= \langle a\rangle$ when $\phi^2 = \langle \phi^2\rangle$. The time dependence of the average scale factor is given by
     \begin{equation}
  \langle a(t) \rangle = a_0 \, \left(   {t\over t_0} \right)^{n+2 \over 3n} \;.
    \end{equation}
 We observe that the oscillating correction in $a(t)$ is suppressed by $\phi^2 / M_{\rm Pl}^2$.   
  An oscillating background generally results in particle production, which we consider next.

    \subsection{Decoupled scalar  production in the inflaton oscillation epoch}

    Consider the free scalar action
    \begin{equation}
    S= \int d^4 x \sqrt{|g|} \left[ -{1\over 2} g^{\mu\nu} \partial_\mu s \, \partial_\nu s -{1\over 2} m_s^2 s^2
    \right] \;,
    \label{free-S}
    \end{equation}
    where $g={\rm det}\, g_{\mu\nu}$, $g_{\mu\nu}={\rm diag} (-1, a^2 , a^2 ,a^2)$ and 
     we have neglected scalar self--interaction.
    In a homogeneous inflaton background, the equation of motion for $s$ reads
    \begin{equation}
    \ddot s - {1\over a^2 } \partial_i \partial_i s + 3H \dot s + m_s^2 s =0\;.
    \end{equation}
    Decomposing $s$ into spacial  comoving  momentum modes $s_k$ and defining  a new variable $s_k = a^{-3/2} X_k$, one finds
     \begin{equation}
    \ddot X_k +  \left[   {9\over 4}wH^2 + {k^2\over a^2} + m_s^2    \right] X_k =0\;,
    \end{equation}
    where $w$ is the coefficient of the equation of state of the system,
    \begin{equation}
    w=- \left(  1+ {2 \dot H \over 3H^2} \right) \;.
    \end{equation}
    Decomposing the Hubble rate and $a$ into the average value and an oscillating correction, we get 
     \begin{equation}
    \ddot X_k +  \left[   {9\over 4}\langle w\rangle \langle H^2\rangle  +{k^2\over \langle a\rangle^2} + m_s^2      -{3\over 2} \delta \dot H -   {2 k^2\over \langle a\rangle^3} \,   \delta a  \right] X_k =0\;,
    \label{EOM-Xk}
    \end{equation}
where we have kept only linear terms in $\delta H, \delta a$ and  omitted $H \delta H \ll \delta \dot H$ as before.\footnote{This equation also explains particle production during inflation: $\langle w\rangle =-1$ implies 
tachyonic amplitude growth for light fields.} 
    Since $\delta H$ and $\delta a$ are (approximately) periodic, this belongs to the class of Hill's equations.
    The solutions can oscillate or grow in time depending on the parameter region.
     The amplitude growth is interpreted as particle production \cite{Kofman:1997yn,Greene:1997fu}.

     The particle production efficiency is determined by two parameters:
     (1) the relative size of the oscillating term in the equation of motion of $X_k$ compared to the inflaton oscillation frequency; (2)
       the relative size of the oscillating and non-oscillating terms. 
       If either of these is small, the resonance is inefficient.   We assume that $m_s$ is far below the energy scales of the relevant processes, so we may omit it.
       At low $k$, the relevant quantity characterising the resonance  is 
        \begin{equation}
     q \sim { \delta \dot H \over m_\phi^{\rm eff\,2}} \sim {\phi^2 \over M_{\rm Pl}^2} \ll 1  \;,
    \end{equation}
     hence there is no significant resonant enhancement of $X_k$. At larger  $k$, the strength of the resonance is controlled by the ratio of the oscillating term to the constant term, 
     \begin{equation}
     {\delta a \over \langle a \rangle}  \sim {\phi^2 \over M_{\rm Pl}^2} \ll 1 \;,
     \end{equation}
     therefore the resonance is again inefficient. Specifically,
     in the quadratic inflaton potential, one gets a Mathieu type equation $X_k^{\prime\prime} + \left[      A_k + 2q_k \cos 4z    \right] X_k=0$, where $z=m_\phi t /2$ and the prime denotes differentiation with respect to $z$.
     One finds $q_k \ll 1$ at low $k$ and $q_k /A_k \ll 1$ at significant $k$. As is clear from the Mathieu equation stability chart, both of these regimes lead to no tangible resonant effects, especially if one takes into account the Universe expansion and 
     the redshifting of the produced particle momenta \cite{Mukhanov:2005sc}. A similar statement applies to the quartic inflaton potential leading to a Lam\'e type equation \cite{Greene:1997fu}. .

          \subsubsection{Perturbative estimate}
     
    We observe that although some particle production takes place during the inflaton oscillation epoch, the associated collective effects are insignificant.  One may therefore estimate the rate of particle production perturbatively \cite{Dolgov:1989us,Traschen:1990sw,Ichikawa:2008ne}.
    The oscillating scale factor induces the  effective couplings in the Lagrangian
    \begin{equation}
    {C\over M_{\rm Pl}^2 } \, \phi^2 (\partial_\mu s)^2  ~~, ~~ {\tilde C \over M_{\rm Pl}^2 } \, \phi^2 (\partial_0 s)^2
    \end{equation}
    according to Eqs.\,\ref{a-osc},\ref{free-S}, where $C\simeq {1\over 4(n+2)} $, $\tilde C\simeq -{1\over 2(n+2)}$.  An analogous coupling $\phi^2 s^2$ is suppressed by $m_s^2 / M_{\rm Pl}^2$ and can be neglected,
    together with the higher order $\phi^n (\partial_\mu s)^2 $ terms. When $\phi$ oscillates, the above couplings result in $s$--pair production.
    
    Let us  compute the scalar abundance due to the $C$--coupling.
    Expanding the inflaton field as
        \begin{equation}
\phi^2(t) = \sum_{n=-\infty}^\infty \zeta_n e^{-in \omega t} \;,
\label{zeta-def}
 \end{equation}
one finds that the $s$--pair production rate per unit volume is     \cite{Lebedev:2022ljz}
         \begin{equation}
 \Gamma =     {C^2 \, \omega^4 \over 16 \pi M_{\rm Pl}^4}   \,  \sum_{n=1}^\infty  n^4 \vert \zeta_n \vert^2 \;,
  \label{Gamma-ss-1}
  \end{equation}
  where the scalar mass has been neglected ($\omega \gg m_s$).
For the quadratic inflaton potential, $\omega = m_\phi$ and  the sum contains a single term $n=2$. For the quartic potential, the frequency is time--dependent, $\omega \simeq 0.9 \sqrt{\lambda_\phi} \Phi$,
where $\Phi$ is the inflaton oscillation amplitude and the sum can be approximated by the leading term with $n=2$ (see e.g.\,\cite{Greene:1997fu,Lebedev:2021xey}).  

As long as the  rate is low enough and the produced particles are dilute, the backreaction effects can be neglected and the $s$--number density is found via the Boltzmann equation
\begin{equation}
\dot n + 3Hn = 2\Gamma \;,
\end{equation}    
    where the factor of 2 accounts for two produced quanta in each reaction. The rate $\Gamma$ is time dependent: for a $\phi^2$ potential, $\omega$ is constant and $\Gamma \propto
 1/a^6   $, whereas in the quartic potential $\omega \propto 1/a$ and $\Gamma \propto 1/a^8$. 
    Integrating the Boltzmann equation, one finds that the number of the produced $s$--quanta is determined entirely by the early time dynamics. Taking $a_0=1$ and $H=H_0$ at the onset of the 
    inflaton oscillation process, at $a\gg 1$ we have 
    \begin{equation}
    n(t)\simeq {c(\phi_0,n) \over a^3 H_0} \;, 
    \end{equation}
    where $c(\phi_0,n)$ is a dim-4 constant depending on the form of the inflaton potential and initial conditions. We observe that most of the $s$--quanta are created within one $e$--fold after inflation,
    while their total number remains constant at later times.

    Let us determine the corresponding $s$--abundance at the reheating stage according to  
     (\ref{Y-def}). For the $\phi^4$ inflaton potential, the abundance is independent of the reheating temperature. Identifying the beginning of the inflaton oscillation epoch with the end of inflation, 
     $H_0 \sim H_{\rm end}$, the constraint (\ref{Y-constraint}) requires
     \begin{equation}
     m_s \lesssim  10^{-6} \, \left( M_{\rm Pl} \over H_{\rm end}\right)^{3/2} \; {\rm GeV} \;.
     \end{equation}
     For $H_{\rm end} \sim 10^{14}$ GeV, this translates into $m_s \lesssim 5$ GeV.  The constraint is independent of $\lambda_s$ as long as it is much smaller than one, in which case the
     $\lambda_s$--induced backreaction can be neglected. One may also verify that the energy density of $s$ is far below that of the inflaton, so our approximation is self--consistent.
    It is important that most of the $s$ quanta are produced at the initial moments of the inflaton oscillation since the background decoheres quite quickly in the $\phi^4$ potential \cite{Khlebnikov:1996mc},
    making our approach inapplicable at later times.
    Finally, the $\tilde C$--induced production of $s$ yields a similar result, so the above constraint  gives a reasonable, conservative estimate.

     In case of the quadratic inflaton potential, the analysis proceeds similarly, except the resulting bound depends on the reheating temperature. It can be put in the form
     \begin{equation}
     m_s \lesssim   10^{-6} \,   \Delta_{\rm NR}   \,\left( M_{\rm Pl} \over H_{\rm end}\right)^{3/2}  \;{\rm GeV} \;,
     \label{Ema-bound}
     \end{equation}
where $\Delta_{\rm NR} = (H_{\rm end}/H_{\rm reh})^{1/2}$ determines the duration of the non--relativistic expansion period, as before. For instant reheating, one recovers approximately 
the $\phi^4$ result. Taking $H_{\rm end} \sim 10^{14}$ GeV as the benchmark value, we get roughly 
 \begin{equation}
     m_s \lesssim {\rm few} \times \Delta_{\rm NR} \;{\rm GeV} \;.
     \label{ms-Ema}
     \end{equation}
     We observe that this bound is weaker than the analogous bound based on scalar production during inflation. Yet, it involves a different production mechanism and as such sets 
     an independent constraint. For example, particle production during inflation can be suppressed 
    via a large  inflaton--induced mass for the scalar, whereas the above considerations would still apply.

      \section{Quantum gravity effects}
      \label{quantum}

    Quantum gravity is believed to induce all operators consistent with gauge symmetry. 
    To understand scalar production after inflation, we may resort to the effective field theory approach and expand the Lagrangian in $M_{\rm Pl}^{-1}$. 
    To be conservative, let us assume that the $\phi \leftrightarrow -\phi$ symmetry is violated very weakly
    such that the leading operators coupling the inflaton to the scalar $s$ are dim-6  \cite{Lebedev:2022ljz},
         \begin{eqnarray}
 \Delta {\cal L}_6 = {C_1 \over M_{\rm Pl}^2 } \; (\partial_\mu \phi)^2  s^2 +   {C_2 \over M_{\rm Pl}^2 } \;  (\phi \partial_\mu \phi) (s \partial^\mu s) +
 {C_3 \over M_{\rm Pl}^2 } \; (\partial_\mu s)^2 \phi ^2 - {C_4 \over M_{\rm Pl}^2 } \; \phi^4 s^2 
 - {C_5 \over M_{\rm Pl}^2 } \; \phi^2 s^4 \;, 
 \label{L}
 \end{eqnarray}
 Such operators lead to particle production during the inflaton oscillation epoch and can cause excessive abundance of stable scalars.

% \subsection{Perturbative regime}

If the couplings are small enough, collective effects in particle production are unimportant and we can estimate the particle abundance perturbatively.
Not all of the above operators are independent when the particles are on--shell: 
  two of the derivative operators can be eliminated via integration by parts:
\begin{eqnarray}
&& (\partial_\mu \phi)^2  s^2  \rightarrow  (\partial_\mu s)^2 \phi ^2 + m_\phi^2 \phi^2 s^2 \;, \\
&& (\phi \partial_\mu \phi) (s \partial^\mu s) \rightarrow - {1\over 2} (\partial_\mu s)^2 \phi^2 \;.
\end{eqnarray}
Here we have neglected $m_s$ and the Hubble rate,  which, during preheating  is  small compared to the particle energy.  
We are mostly interested in scalar pair production, so we may restrict ourselves to the operators 
\begin{equation}
{\cal O}_3 =  {1 \over M_{\rm Pl}^2 } \; (\partial_\mu s)^2 \phi ^2 ~~,~~ {\cal O}_4= {1 \over M_{\rm Pl}^2 } \; \phi^4 s^2 \;,
\end{equation}
amended with the renormalizable interaction
\begin{equation}
{\cal O}_{\rm renorm} =  {m_\phi^2 \over M_{\rm Pl}^2 } \;  \phi^2 s^2 \;.
\label{O-ren}
\end{equation}
Although this term is renormalizable, the corresponding coupling  is   suppressed: for typical inflaton masses, it is  below ${}10^{-10}$.

{\underline {{\bf Operator} $\bf {{\cal O}_3}$}}.
The effect of ${\cal O}_3 $ was considered in the previous section. Trivially rescaling the result, we find that 
the $s$ quanta are overproduced unless
  \begin{equation}
     |C_3| \lesssim  5\times  10^{-5} \,   \Delta_{\rm NR}^{1/2}   \,\left( M_{\rm Pl} \over H_{\rm end}\right)^{3/4}  \;\sqrt{{\rm GeV} \over m_s }\;,
     \label{c3}
     \end{equation}
 where $ \Delta_{\rm NR}=1$ for the quartic inflaton potential and 
  $\Delta_{\rm NR} = (H_{\rm end}/H_{\rm reh})^{1/2}$ for the quadratic one. Assuming a typical $H_{\rm end} \sim 10^{14}$ GeV, we have 
  \begin{equation}
     |C_3| \lesssim 10^{-1}  \, \Delta_{\rm NR}^{1/2}  \;\sqrt{{\rm GeV} \over m_s }\;,
     \end{equation}
which imposes a non--trivial constraint for $m_s \gtrsim 1$ GeV.

{\underline {{\bf Operator} $\bf {{\cal O}_4}$}}.
The operator ${\cal O}_4$ is much more efficient in particle production than ${\cal O}_3 $ is: the energy factor $\partial \sim m_\phi^{\rm eff}$ is replaced by the inflaton field value $\phi$. 
The reaction rate per unit volume is given by \cite{Lebedev:2022ljz}
 \begin{equation}
 \Gamma =   { C_4^2 \over 4\pi M_{\rm Pl}^4 } \, \sum_{n=1}^\infty  |\hat\zeta_n|^2   \;,
 \label{Gamma-ss-2}
  \end{equation}
where
\begin{equation}
\hat \zeta_n = \sum_{m=-\infty}^\infty \zeta_{n-m} \zeta_m 
 \end{equation}
and $\zeta_n$ is defined by Eq.\,\ref{zeta-def}. 
The rate drops very fast with time: $\Gamma \propto 1/a^8$
 for the quartic inflaton potential and $\Gamma \propto 1/a^{12}$ for the quadratic one. Thus, the particle number is determined entirely  by the early time dynamics. 
 Solving $\dot n + 3Hn = 2\Gamma $ as before, one finds that the $s$--abundance does not exceed that of dark matter for 
 \begin{equation}
 |C_4| < 10^{-3} \, \Delta_{\rm NR}^{1/2} \;     {H_{\rm end}^{5/4} \, M_{\rm Pl}^{11/4} \over \phi_0^4 }  \;\sqrt{{\rm GeV} \over m_s }\;,
 \label{c4-bound}
 \end{equation}
 where $\phi_0$ is the inflaton value at the beginning of the inflaton oscillation epoch, which we identify for simplicity
 with the end of inflation, 
   $H_0  \sim H_{\rm end}$. To appreciate the strength of this constraint, let us assume the typical values $\phi_0 \sim M_{\rm Pl}$ and 
$H_{\rm end} \sim 10^{14}$ GeV. Then   
\begin{equation}
 |C_4| < {\rm few} \times 10^{-9} \, \Delta_{\rm NR}^{1/2} \;    \sqrt{{\rm GeV} \over m_s }\;,
 \label{strong-bound}
 \end{equation}
 which imposes a very strong constraint in case of the quartic inflaton potential or almost instant reheating, $\Delta_{\rm NR} \sim 1$. For $m_s \gtrsim 1$ GeV,
 the size of $C_4$ is constrained at the level $10^{-9}$. Even if one takes the extreme case of an MeV--scale reheating temperature, i.e. $\Delta_{\rm NR} \sim 10^{18}$ 
 for the $\phi^2$ potential, the constraint remains non-trivial as long as  $m_s $ is above the GeV scale.

Clearly, it is hard to imagine how quantum gravity effects could be controlled with such accuracy, unless a UV--complete theory is available.
It is important  to note that this bound comes from inflaton dynamics immediately (within around 1 e-fold) after inflation and is therefore insensitive to 
late time effects such as reheating mechanisms, inflaton decoherence, etc.

 {\underline {{\bf Higher dimensional operators}}}.
When the initial inflaton value at preheating is not far below the Planck scale, operators of dimension higher than 6 are also important. Consider a dimension $p+2$ operator
\begin{equation}
{\cal O}^{(p)}= { \phi^p s^2 \over M_{\rm Pl}^{p-2}} \;.
\end{equation}
Repeating the above calculations, one finds the following constraint on the corresponding Wilson coefficient:
\begin{equation}
 |C^{(p)}| < 10^{-3} \, \Delta_{\rm NR}^{1/2} \;     {H_{\rm end}^{5/4} \, M_{\rm Pl}^{p-5/4} \over \phi_0^p }  \;\sqrt{{\rm GeV} \over m_s }\;,
 \end{equation}
where we have omitted a mild $p$--dependence of the prefactor. Compared to the $p=4$ case, the bound is weaker by the factor $(M_{\rm Pl}/\phi_0)^{p-4}$.
Given the strength of the bound (\ref{strong-bound}), the resulting constraints can be significant even for high $p$, for instance, 8 or 10.

 {\underline {{\bf Renormalizable coupling}}}.
In general,  a renormalizable inflaton--scalar coupling
\begin{equation}
V_{\phi s} = {1\over 4} \lambda_{\phi s} \phi^2 s^2
\end{equation}
is present. Naturally, it leads to intense particle production unless the coupling is tiny.
The scalar abundance  constraint on $ \lambda_{\phi s}$
has the form 
\begin{equation}
|  \lambda_{\phi s} | < 10^{-3} \, \Delta_{\rm NR}^{{\pm}1/2} \;     {H_{\rm end}^{5/4} \, M_{\rm Pl}^{3/4} \over \phi_0^2 }  \;\sqrt{{\rm GeV} \over m_s }\;,
 \end{equation}
where $\Delta_{\rm NR} = (H_{\rm end}/H_{\rm reh})^{1/2}$, the plus sign applies to the $\phi^2$ local inflaton potential and the minus sign applies to the $\phi^4$ potential.

Note that, unlike before, $\Delta_{\rm NR}$ dependence appears also in the quartic inflaton potential case. This is due to the 
fact that the 
particle number grows linearly with $a$ and the result is sensitive to the late time behaviour of the system. Clearly, in this case
 $\Delta_{\rm NR}$ does not represent the duration of the non--relativistic expansion period, but rather the duration of the coherent inflaton oscillation phase, during which 
 the $s$--quanta are produced. In reality, the inflaton background often loses coherence before reheating since the $\phi^4$ potential leads to inflaton quanta production via parametric 
 resonance \cite{Khlebnikov:1996mc}. Therefore,  $\Delta_{\rm NR}$ cannot be too large in this case, although the exact numbers are model dependent. To illustrate our point, it suffices to take 
$\Delta_{\rm NR}\sim 1$ for  the quartic potential, which already imposes a strong bound on $\lambda_{\phi s}$. One should keep in mind, however, that the bound can be significantly stronger
depending on further details, e.g. the relative size of the inflaton self--coupling and $\lambda_{\phi s}$.

Using $\phi_0 \sim M_{\rm Pl}$ and 
$H_{\rm end} \sim 10^{14}$ GeV as the benchmark values, we get 
 \begin{equation}
|  \lambda_{\phi s} | < 10^{-8} \, \Delta_{\rm NR}^{\pm 1/2}      \;\sqrt{{\rm GeV} \over m_s }\;.
\label{lambda-phi-s}
 \end{equation}

For larger couplings,  collective effects become significant and particle production becomes even more efficient. These were studied in Ref.\,\cite{Lebedev:2022ljz},\cite{Lebedev:2021tas}
in the case of  dim-4 and dim-6 operators.
For  operators of the form $\phi^p s^2$ and a quadratic inflaton potential, resonant particle production is described by the semi-classical EOM \cite{Dufaux:2006ee}
\begin{equation}
\ddot X_k + \left[    \alpha + \beta \cos^p m_\phi t    \right] X_k=0 \;,
\end{equation}
where $X_k$ is the Fourier mode of the scalar field $s$ with 3-momentum $k$, and $\alpha,\beta$ are slowly varying coefficients. This is an analog of the Mathieu equation,
while in the quartic case, one gets an analog of the Lam\'e equation. At large enough couplings,
$\beta$ is significant and leads to resonant amplification of the amplitude, which is interpreted as particle production. To include backreaction effects as well as rescattering of the produced quanta, it is however necessary to perform lattice simulations.
This was done in \cite{Lebedev:2022ljz},\cite{Lebedev:2021tas}. For typical inflaton masses, such collective effects are important when $\lambda_{\phi s} > 10^{-8}$ or $C_4 > 10^{-7}$.\footnote{In the
 $\phi^4$ potential, resonant effects are important for $\lambda_{\phi s} \gtrsim \lambda_\phi$, where $\lambda_\phi$ is the inflaton self-coupling.}
For our purposes, however, it is sufficient to restrict ourselves to the perturbative regime, in which case the resulting bounds are already strong.

Finally, the effect of operator $\phi^2 s^4 $ is not as significant as that of $\phi^4 s^2$, due to the lower power of the inflaton field  \cite{Lebedev:2022ljz}.
 It leads to a constraint analogous to (\ref{c3}) and will not be considered separately. The dim-5 operator $\phi^3 s^2$, on the other hand, 
  can be the dominant source of scalar production. However, it is odd under $\phi \rightarrow - \phi$, so one may suppress its coefficient using symmetry arguments, at least in principle.
  To be conservative, we will simply assume that its effect is small.

We conclude that quantum gravity effects in the form of higher dimensional Planck--suppressed operators generally lead to efficient scalar production in the early Universe.
The resulting scalar abundance often exceeds that of dark matter, which puts strong constraints on the UV complete models of gravity. 
These bounds are evaded if the stable scalars are very light, far below the GeV scale, or the Universe undergoes an extremely  long period of non--relativistic expansion, corresponding to very low reheating temperatures.

  \section{Freeze-in production}  
  
  Another significant source of scalar production is the coupling between $s$ and the Standard Model fields. In particular, after reheating, its interaction with the Higgs field,
  \begin{equation}
  V_{hs}= {1\over 2} \lambda_{hs} H^\dagger H s^2 \;
  \end{equation}
  leads to thermal emission of the $s$--quanta. This is known as ``freeze-in'' production    \cite{McDonald:2001vt,Kusenko:2006rh,Hall:2009bx,Yaguna:2011qn}   since the scalar abundance accumulates
  gradually   until the thermal bath becomes devoid of the Higgs bosons.
  This mechanism is very efficient as long as the temperature is above the Higgs and the scalar masses, $T > m_h, m_s$. The resulting constraint on the Higgs portal coupling  reads \cite{Lebedev:2019ton}
   \begin{eqnarray}
  &&\lambda_{hs} < 2 \times 10^{-11}            ~~~~~~~~ {\rm for ~~} m_s \gtrsim m_h  \;,    \nonumber \\
     &&\lambda_{hs} <   10^{-11} \; \sqrt{ {\rm GeV} \over m_s}    ~~ ~{\rm for ~~}         m_s \ll m_h\;.
  \end{eqnarray}
  The freeze-in approximation is adequate if the scalar does not thermalize, which requires $\lambda_{hs} \ll 10^{-7} \sqrt{m_s / {\rm GeV}}$.
  
  Clearly, the constraint is strong and the Higgs portal coupling above $10^{-11}$ must be forbidden unless the scalar is at the GeV scale or below.
  Although such a tiny coupling is $technically$ natural in the sense that radiative corrections are proportional to the coupling itself, it is a highly non--trivial task to build a 
  UV complete model where  $\lambda_{hs} <  10^{-11} $ would appear naturally. 
  
  The quantum gravity effects on the freeze-production are less significant than those considered in the previous section. Indeed, operators of the type $H^\dagger H s^4 /M_{\rm Pl}^2 $ and alike are suppressed by
  the ratio of the relevant energy scale and the Planck scale. The resulting reaction rate exhibits a suppression factor  $T^4/ M_{\rm Pl}^4$ which makes it insignificant unless the temperature is very high.
  Although this may lead to non--trivial constraints, such effects are not essential for our purposes.

  Finally, let us note that the $s$--quanta can also be produced by the perturbative inflaton decay $\phi \rightarrow ss$. This is a well-known mechanism, which we will not discuss further in this work.
   A relevant analysis can be found, for example, in Ref.\,\cite{Lebedev:2021xey} (Section 8.4).

  \section{Implications and discussion}

 The existence of very weakly interacting stable particles imposes significant constraints on cosmological models with high scale inflation. These  are produced at various stages of the Universe evolution which can cause 
 overabundance of  dark relics. 
 
In fact,  such constraints equally apply to
 very long-lived particles. If their abundance is similar to that of dark matter, their lifetime is required to be much longer than the age of the Universe, $\tau \gtrsim 10^{25}\;$sec, otherwise
 the CMB spectrum gets significantly distorted \cite{Slatyer:2016qyl}. In case of a smaller abundance of dark relics, 
 the energy injection into the CMB   can still be significant  such that the Early Universe particle production would still impose non-trivial constraints. This aspect deserves a separate study.

  Below we discuss some of the implications of stable particle production in the Early Universe.

  \subsection{Quantum gravity effects on  non-thermal dark matter}
  \label{non-th-dm}
  
  The above considerations have  implications for non-thermal dark matter. As illustrated in Fig.\,\ref{fig-intro}, its abundance is determined by contributions of a number of  sources active at 
  different stages of the Universe evolution. One of them can be  attributed to  quantum gravity in the form of higher dimensional Planck--suppressed operators. In particular, non-derivative operators of the type
      \begin{equation}
  { \phi^4 s^2 \over M_{\rm Pl}^{2}} ~~,~~  { \phi^6 s^2 \over M_{\rm Pl}^{4}} ~~,~~  { \phi^8 s^2 \over M_{\rm Pl}^{6}}~~,...
\end{equation}
are efficient sources of dark matter during the inflaton oscillation epoch, which is (almost) always present in standard cosmology. Since the corresponding rates fall off fast with the scale factor,
contributions of these operators to the final DM abundance are dominated by the initial moments after inflation. This makes such effects insensitive to the   details of the reheating dynamics.
  Since the inflaton field value is not far from the Planck scale at the end of inflation, the above operators are almost unsuppressed and can readily  overproduce dark matter.

  Unless dark matter is very light, far below the GeV scale, or the Universe undergoes a very long non-relativistic expansion period, the constraints on the Wilson coefficients of such operators
  are very strong, at the level of $10^{-8}$  (cf.~Eq.\,\ref{strong-bound}). Therefore, in order to compute relic abundance of non-thermal dark matter, one needs to control quantum gravity effects with high accuracy.
  Clearly, this is only possible within a UV complete theory of gravity, unless the above operators are forbidden by $gauge$ symmetry.

 In order to suppress the above operators, one may want to invoke the inflaton shift symmetry $\phi \rightarrow \phi + {\rm const}$, which is often quoted as the argument for the flatness of its potential.
 However, the effects discussed here occur around the minimum of the inflaton potential, where this symmetry is violated. Hence, this approach does not appear fruitful.
 
 Similar considerations apply to fermionic dark matter $\Psi$. For example, an operator
   \begin{equation}
  { \phi^2 \, \bar \Psi \Psi \over M_{\rm Pl}}  
\end{equation}
     is expected to be induced by quantum gravity. The rate calculation is analogous to that for $\lambda_{\phi s} \phi^2 s^2$ as long as collective effects are insignificant.
     Then the constraint (\ref{lambda-phi-s}) applies, up to the replacement $\lambda_{\phi s} \rightarrow  C_{\rm ferm} m_\phi^{\rm eff}/M_{\rm Pl}$, where $C_{\rm ferm}$ is the corresponding Wilson coefficient.
     The resulting constraint on $C_{\rm ferm}$ is non--trivial, at the level of $10^{-2}-10^{-3}$ for GeV masses, meaning that fermionic dark matter is also copiously produced after inflation.
   
  It is worth noting  that semi-classical gravity can have a non-negligible impact on the relic abundance due to graviton-mediated processes active at very high reheating temperatures \cite{Mambrini:2021zpp}.
   Such processes can be mimicked by dim-6 Planck-suppressed operators of the ${\cal O}_3$ type, which  generate an $m_\phi^4/M_{\rm Pl}^4$ suppressed contribution.

  We  conclude that the abundance of non-thermal dark matter is very sensitive to quantum gravity effects. These spoil predictivity of non-thermal dark matter models unless 
  DM is very light or the Universe undergoes a very long non-relativistic expansion period resulting in a low reheating temperature.

  \subsection{Suppressing inflationary particle production}
  
  Depending on the size of the scalar--inflaton couplings, particle production during inflation can be suppressed. This occurs when the scalar becomes heavy due to the effective inflaton--induced mass. 
  Assuming one coupling at a time, if 
  \begin{equation}
  \sqrt{\lambda_{\phi s }}\, \phi ~,~ \sqrt{C_4} \,  {\phi^2 \over M_{\rm Pl}}, ... ~  \gtrsim ~ H 
  \label{suppress}
  \end{equation}
  during inflation, fluctuations of the scalar field are suppressed and  its abundance at the end of inflation can be neglected.\footnote{Here we assume positive couplings to avoid complications with $s$ acquiring a VEV.}  The  consequent bounds on the couplings for the lowest order operators read $\lambda_{\phi s} \gtrsim m_\phi^{\rm eff\, 2}/ M_{\rm Pl}^2 \;,\; C_4 \gtrsim m_\phi^{\rm eff\, 2}/ \phi^2$, which
  for  typical inflaton masses are of order $10^{-10}$. We note that it is sufficient to require the above inequality to apply at the end of inflation, because $H$ grows slower than $\phi$ with the inflaton field value.

  A large induced scalar mass  effectively switches off inflationary particle production, however postinflationary scalar production remains (or becomes more) efficient.
  In this case, only the bounds like (\ref{strong-bound}) and (\ref{lambda-phi-s}) apply. If the above inequality  is violated, however, the scalar abundance receives both contributions
  such that   (\ref{ms-infl}) or (\ref{m-lambda-NR})   must also be satisfied. While for extremely light scalars, far below  the eV scale, such bounds are easily observed,  generally the constraints are quite strict.
  
  Furthermore, there is always particle production induced by an oscillating component of $H$ after inflation, as discussed in Section
 \ref{Ema-effect}.  This corresponds to an effective coupling $C_3 \sim 10^{-1}$ generated by classical gravity. Thus, the bounds on the Wilson coefficients are to be supplemented with 
 the constraint (\ref{Ema-bound}). For $H_{\rm } \sim 10^{14}$ GeV, this amounts to 
 \begin{equation}
 \left(      {\Delta_{\rm NR} \over m_s/{\rm GeV}}    \right)^{1/2} \gtrsim 1 \;.
 \end{equation}  
  Therefore, even if inflationary scalar production is suppressed, postinflationary dynamics set strong bounds on stable scalars.

  \subsection{Implications for models of inflation: an example}
  
  The particle production constraints are sensitive to the details of the inflation model, in particular, the dilution factor $\Delta_{\rm NR}$.
  In what follows, we illustrate the strength of the constraints with an example.

  Let us consider a popular inflationary model, where expansion  is  driven by a non--minimal scalar coupling to curvature akin to {\it Higgs inflation} \cite{Bezrukov:2007ep}.    
  In this case, postinflationary dynamics takes place in two regimes corresponding to the quadratic and quartic local inflaton potentials, 
   which requires a more careful treatment of  $\Delta_{\rm NR}$.

  Consider the action based on \cite{Bezrukov:2007ep}
  \begin{equation}
{\cal L}_{J} = \sqrt{-\hat g} \left(   -{1\over 2}  \Omega  \hat R \,  
 +  {1\over 2 } \, \partial_\mu \phi \partial^\mu \phi   - {V(\phi)  }\right) \;,
\label{L-J}
\end{equation}
where we use the Planck units $M_{\rm Pl}=1$,
$\hat g_{\mu \nu}$ is the Jordan frame metric,  $\hat R$ is the scalar curvature, $V(\phi) \simeq {1\over 4} \lambda_\phi \phi^4$, and 
  \begin{equation}
 \Omega = 1 + \xi_\phi \phi^2 \;,
 \end{equation} 
 with constant $\xi_\phi$.
  Making a conformal transformation to the Einstein frame, $g_{\mu \nu}= \Omega \hat g_{\mu\nu}$, one finds that the canonically normalized inflaton $\chi$ satisfies
  \begin{equation}
{d \chi \over d \phi} = \sqrt{  1 + \xi_\phi (1+6 \xi_\phi) \phi^2  \over  (1 +\xi_\phi \phi^2)^2 } \;.
\label{dchi/dphi}
 \end{equation}
  For $6\xi_\phi \gg 1$, at large field values one finds the inflationary potential $V_E(\chi)= {\lambda_\phi \over 4 \xi_\phi^2 }\, \left(     1- e^{ \sqrt{2\over 3} |\chi |} \right)^2$, while for smaller $\chi$
  after inflation we have \cite{GarciaBellido:2008ab}
\begin{equation}
V_E(\chi) \simeq \left \{
  \begin{tabular}{ccc}
 ${\lambda_\phi \over 4}   \chi^4$  &  for & $|\chi |\ll {1\over 2 \xi_\phi } ~,$ \\
  $ {\lambda_\phi \over 6 \xi_\phi^2 } \chi^2 $ & for   & ${1\over 2 \xi_\phi } \ll |\chi | \ll  1  ~,$
  \end{tabular}
\right. 
\end{equation}   
in Planck units.
Therefore, after inflation, the Universe expands in a quadratic potential. Most significant particle production  occurs at the end of inflation,
if collective effects are small, while after that the total particle number remains approximately constant. 
Later, at  $a/a_0 \sim \xi_\phi^{2/3}$,
 the system starts scaling as radiation.  In this case, 
  the dilution factor $\Delta_{\rm NR}$ is  defined as $(H_{\rm end}/H_{\rm rad})^{1/2}$, where $H_{\rm rad}$ signifies 
  the onset  of radiation domination rather than reheating.
  We thus have 
\begin{equation}
\Delta_{\rm NR} \sim \xi_\phi^{1/2} \;.
\end{equation} 
 In this model, 
 the CMB normalization does not allow for a very large 
 $\xi_\phi$, above a few hundreds  \cite{Ema:2017ckf}, without violating unitarity \cite{Burgess:2009ea,Barbon:2009ya,Ema:2016dny}. Thus, the non--relativistic expansion period is not particularly long.
 Here we assume, as usual, that the bare inflaton mass is small enough such that reheating occurs before the inflaton becomes non--relativistic.

 In this example, the dilution factor $\Delta_{\rm NR}$ represents the duration of the non--relativistic expansion period and is not immediately related to the reheating temperature.
 Since it is bounded from above by ${\cal O}(10)$ and $H_{\rm end} \sim 10^{13}\;$GeV, the constraints on feebly interacting stable scalars in this model are quite strong. According to (\ref{Ema-bound}), such scalars with masses above 1 TeV or so
 are ruled out. The inflationary constraint depends on the size of the $\phi-s$ couplings. If these are very small, the bound (\ref{m-lambda-NR}) yields\footnote{Depending on $\xi_\phi$ and $\lambda_s$,
the $s$-condensate may start oscillating before or after the transition to the radiation-like scaling. The difference between the two possibilities is not fundamental, hence we simply use  (\ref{m-lambda-NR})
which assumes the first option.}
 \begin{equation}
 m_s \lesssim 10^{-1} \; {\rm GeV} \;,
 \end{equation}
 taking into account  the non-thermalization constraint on $\lambda_s$ \cite{Arcadi:2019oxh}.
 On the other hand, if the $\phi-s$ couplings are substantial as in (\ref{suppress}), the inflationary constraint is not applicable and we only get a set of constraints on
 $\lambda_{\phi s}$ and Planck--suppressed operators listed in Section \ref{quantum}. For $\Delta_{\rm NR} \lesssim 10$, they are all very strong.
 Generic order one Wilson coefficients are allowed only if (cf.\,Eq.\,\ref{c4-bound})
 \begin{equation}
 m_s \lesssim 10^{-16} \, {\rm GeV} \;.
 \end{equation}
  We thus conclude that the existence of stable feebly interacting scalars in this framework would be problematic, unless they are extremely light or quantum gravity corrections are well under control.

  In more general models, the factor $\Delta_{\rm NR} $ can be very large. Consider a locally quadratic inflaton potential together with the reheating mechanism driven by perturbative inflaton
  decay into the Higgs pairs. The Higgs--inflaton coupling is given by 
  \begin{equation}
  V_{\phi h} =   \sigma_{\phi h} \, \phi H^\dagger H \;.
 \end{equation}
  Taking into account 4 Higgs d.o.f. at high energies, we have
   \begin{equation}
  \Delta_{\rm NR} \simeq \left(   {H_{\rm    end}  \over \Gamma_\phi}      \right)^{1/2} \sim {m_\phi \over \sigma_{\phi h}} \, \left(   {\phi_0  \over M_{\rm Pl}}      \right)^{1/2} \;.
   \end{equation}
For small $ \sigma_{\phi h}$, the dilution factor can be huge. The size of $ \sigma_{\phi h}$ is only restricted by the lower bound on the reheating temperature, which is around a few MeV \cite{Hannestad:2004px}.
In the extreme case, $\Delta_{\rm NR}$ can be as large as $10^{18}$. Most of the bounds on particle production become then irrelevant. However, the constraints on Planck-suppressed operators are still 
meaningful: for example, $|C_4| \lesssim \sqrt{{\rm GeV} \over m_s}$ sets an important constraint on quantum gravity for $m_s \gg 1$\, GeV. Similarly, bounds on the higher dimensional operators
$\phi^p s^2/ M_{\rm Pl}^{p-2}$ remain significant.

\section{Conclusion}

We have studied production of very weakly coupled stable scalars in the Early Universe. Particle production is most intense during and immediately after  (high scale) inflation, which can readily lead to overabundance of dark relics. The observed value of the dark matter density sets strong upper bounds on the mass scale of stable scalars, which depend on the duration of the non-relativistic expansion period
in the Universe evolution. In typical inflationary  models akin to Higgs inflation, such bounds are in the sub-GeV range.

Quantum gravity effects in the form of higher dimensional Planck--suppressed operators make an important impact on the abundance of stable scalars. Operators of the type
$\phi^p s^2/ M_{\rm Pl}^{p-2}$ lead to intense particle production immediately after inflation, as long as the inflaton field value is not far from the Planck scale.
Unless Wilson coefficients of such operators are very small, this  mechanism can overproduce dark relics.
This implies, in particular, that in models where the dark state abundance receives additive contributions at various stages of the Universe evolution,
the relic density calculation is  marred by quantum gravity.  
In the absence of full control of quantum gravity effects,  predictivity of non--thermal dark matter models is thus brought into question. The exceptions from such uncertainties are 
models with ultra-light dark matter and  those with an ultra-long  non-relativistic expansion period, corresponding to a very low reheating temperature. Finally, in more complicated models,
one can imagine  suppressing the unwanted operators  by gauge symmetry arguments. 

Throughout this work, we have assumed high scale and large field inflation. The constraints relax significantly in low scale inflationary models.

The above considerations may  have non-trivial implications for string theory constructions (see, e.g.\,\cite{Dumitru:2022eri}), where scalar fields  are ubiquitous.  
\\ \ \\
{\bf Acknowledgements.} The author is indebted to Timofey Solomko and Jong-Hyun Yoon for their invaluable help. Useful comments from Syksy Rasanen are also acknowledged.

\end{document}